\documentclass[conference]{IEEEtran}
\IEEEoverridecommandlockouts

\usepackage{cite}
\usepackage{dblfloatfix}
\usepackage{overpic}
\usepackage{amsmath,amssymb,amsfonts}
\usepackage{graphicx}
\usepackage{textcomp}
\usepackage{xcolor}
\usepackage{float}
\usepackage{amsthm}
\usepackage{graphicx}
\usepackage{epstopdf}
\usepackage{amsmath,bm}
\usepackage{amsfonts}
\usepackage{amssymb}
\usepackage{color}
\usepackage{multirow}
\usepackage{multicol}
\usepackage{soul,xcolor}
\usepackage{algorithm}% http://ctan.org/pkg/algorithms
\usepackage{algpseudocode}

\usepackage{comment}

\theoremstyle{plain}
\newtheorem{theorem}{Theorem}

\newcommand{\vect}[1]{\mathbf{#1}}

\def\diag{\mathrm{diag}}

\def\tr{\mathrm{tr}}

\def\Htran{\mbox{\tiny $\mathrm{H}$}}
\def\Ttran{\mbox{\tiny $\mathrm{T}$}}
\def\CN{\mathcal{N}_{\mathbb{C}}} 
\def\imagunit{\mathsf{j}}

\def\sinc{\mathrm{sinc}}

\def\imaginary{\mathsf{j}} % Imaginary number
\def\imagunit{\imaginary} % Alternative macro

\def\sinc{\mathrm{sinc}}
\def\tr{\mathrm{tr}}

\def\pb{\textrm{p}}

\begin{document}

\title{Wideband Channel Capacity Maximization With Beyond Diagonal RIS Reflection Matrices}

\author{\"Ozlem Tu\u{g}fe Demir, \emph{Member, IEEE}, Emil Bj{\"o}rnson, \emph{Fellow, IEEE} \vspace{-2cm}
\thanks{ \"O. T. Demir is with the Department of Electrical-Electronics Engineering, TOBB University of Economics and Technology, Ankara, T\"urkiye (ozlemtugfedemir@etu.edu.tr).  E.~Bj\"ornson is with the KTH Royal Institute of Technology, 16440 Kista, Sweden (emilbjo@kth.se).  \newline \indent
The work by \"O. T. Demir was supported by 2232-B International Fellowship for Early Stage Researchers Programme funded by the Scientific and Technological Research Council of T\"urkiye. E.~Bj\"ornson was supported by the  FFL18-0277 grant from the Swedish Foundation for Strategic Research.
}\vspace{-3cm}

}

\maketitle
\begin{abstract}
Following the promising beamforming gains offered by reconfigurable intelligent surfaces (RISs), a new hardware architecture, known as \emph{beyond diagonal RIS (BD-RIS)}, has recently been proposed. This architecture enables controllable signal flows between the RIS elements, thereby providing greater design flexibility. However, the physics-imposed symmetry and orthogonality conditions on the non-diagonal reflection matrix make the design challenging. In this letter, we analyze how a BD-RIS can improve a wideband channel, starting from fundamental principles and deriving the capacity. Our analysis considers the effects of various channel taps and their frequency-domain characteristics. We introduce a new algorithm designed to optimize the configuration of the BD-RIS to maximize wideband capacity. The proposed algorithm has better performance than the benchmarks. A BD-RIS is beneficial compared to a conventional RIS in the absence of static path or when the Rician $\kappa$-factor is smaller than $10$.
\end{abstract}

\section{Introduction}

Reconfigurable intelligent surfaces (RISs) have attracted much attention for their low-cost passive beamforming capabilities that can be used to reflect signals from a transmitter to a receiver in challenging environments \cite{10316535}. Many studies have explored passive RISs, where each element independently adjusts the phase shift of the impinging waveform before reflection \cite{yang2020intelligent,sun20223d}. A detailed analysis of RIS-aided narrowband and wideband communication systems is provided in \cite{9721205}.

The conventional RIS architecture features RIS elements individually connected to the ground, enabling independent phase-shift adjustments. This is represented by a diagonal reflection matrix in the system models, where each diagonal entry indicates how the signal impinging on an element is modified before re-radiated by the same element \cite{10316535}. It has recently been emphasized that one can also build \emph{beyond-diagonal RIS (BD-RS)}, allowing interactions between the elements through passive circuitry. Shen et al. \cite{shen2021modeling} categorize RIS architectures based on their internal connectivity, with conventional RIS being single-connected. In contrast, fully- and group-connected RIS networks enable non-diagonal reflection matrices for increased flexibility, as they allow controllable signal flows between RIS elements before reflection \cite{zhou2023optimizing}. This enables power gains in specific scenarios \cite{shen2021modeling,li2022beyond,nerini2023closed,zhou2023optimizing}, particularly when the elements experience different channel conditions.

Wideband communications with the assistance of a BD-RIS have not been previously considered, apart from the concurrent work \cite{li2024wideband}. That paper assumes the RIS phase-shifts vary linearly over the subcarriers and proposes a quasi-Newton method to optimize the channel capacity. 
However, there is no derivation of the proposed system model, which is conjectured to represent orthogonal frequency division multiplexing (OFDM), and no algorithm tailored to the problem structure.
In this paper, we fill these gaps by deriving the OFDM system model from first principles, resulting in a different expression than in \cite{li2024wideband}. Furthermore, we propose a tailored algorithm to optimize the wideband capacity under specific constraints. The resulting capacity expression provides detailed insights into how the propagation paths enter into the cascaded channel expressions. The simulations show that employing BD-RIS, as opposed to conventional RIS, significantly enhances wideband capacity, especially in non-line-of-sight (NLOS) channels.

\section{Derivation of the Wideband System Model}

In this section, we derive the wideband OFDM system model for the operation of a BD-RIS from first principles, following a similar methodology as in \cite[Sec.~9.3]{bjornson2024introduction}.
We consider a communication scenario with a single-antenna transmitter and a single-antenna receiver communicating over a wideband channel, with the assistance of a BD-RIS with $N$ elements. This RIS is implemented using the fully-connected circuit networking topology proposed in \cite{shen2021modeling}, but it is straightforward to extend the work to the group-connected architecture.

We denote the continuous-time passband signal emitted by the transmitter as $x_{\pb}(t)$ and let  $g_{\textrm{s},\pb}(t)$ denote the impulse response of the static linear time-invariant channel in the absence of the RIS. When the signal $x_{\pb}(t)$ reaches a particular RIS element, it passes through the internal fully-connected circuitry and is partially re-radiated from all elements. 
We let $\vartheta_{n,m,\pb; \vect{\Psi}}(t)$ denote the impulse response of the channel between the input of element $n$ and the output of element $m$ through the internal circuit. Additionally, $g_{\textrm{t},m,\pb}(t)$ and $g_{\textrm{r},n,\pb}(t)$ denote the impulse responses of the channels from the transmitter to element $n$ and from that element to the receiver, respectively. Therefore, the received signal (neglecting noise for brevity) can be expressed as
\begin{align} \nonumber
\upsilon_{\pb}(t) &= (g_{\textrm{s},\pb} *x_{\pb})(t) \nonumber\\
&\quad+ \sum_{n=1}^{N} \sum_{m=1}^N(g_{\textrm{r},n,\pb} * \vartheta_{n,m,\pb;\vect{\Psi}} * g_{\textrm{t},m,\pb} *x_{\pb})(t)  \nonumber
\\ &= 
\left(
g_{\pb;\vect{\Psi}}
* x_{\pb} \right)(t), \label{eq:input-output-passband-RIS}
\end{align}
where the end-to-end impulse response of the system is
\begin{equation}
g_{\pb;\vect{\Psi}}(t) = g_{\textrm{s},\pb}(t) + \sum_{n=1}^{N}\sum_{m=1}^M (g_{\textrm{r},n,\pb} * \vartheta_{n,m,\pb; \vect{\Psi}} * g_{\textrm{t},m,\pb} ) (t). \label{eq:end-to-end-impulse-passband}
\end{equation}
We can obtain the complex baseband representation 
\begin{equation} \label{eq:end-to-end-impulse}
g_{\vect{\Psi}}(t) = g_{\textrm{s}}(t) + \sum_{n=1}^{N}\sum_{m=1}^N (g_{\textrm{r},n} * \vartheta_{n,m; \vect{\Psi}} * g_{\textrm{t},m} ) (t)
\end{equation}
of $g_{\pb;\vect{\Psi}}(t)$ in \eqref{eq:end-to-end-impulse-passband} by down-shifting all the impulse responses: 
$g_{\textrm{s}}(t) = g_{\textrm{s},\pb}(t) e^{-\imagunit 2\pi f_{\mathrm{c}} t}$, 
$g_{\textrm{t},m}(t) = g_{\textrm{t},m,\pb} (t)e^{-\imagunit 2\pi f_{\mathrm{c}} t}$, 
$g_{\textrm{r},n} (t) = g_{\textrm{r},n,\pb} (t) e^{-\imagunit 2\pi f_{\mathrm{c}} t}$, 
$\vartheta_{n,m; \vect{\Psi}}(t) = \vartheta_{n,m,\pb; \vect{\Psi}} (t) e^{-\imagunit 2\pi f_{\mathrm{c}} t}$.

The continuous-time complex baseband equivalent of the transmitted passband signal $x_{\pb}(t)$ is denoted by $x(t)$ and it is obtained via pulse-amplitude modulation (PAM) as
\begin{align}
    x(t) = \sum_{k=-\infty}^{\infty}x[k]p\left(t-\frac{k}{B}\right)
\end{align}
where $\{x[k]\}$ is the discrete sequence of data symbols at the symbol rate $B$ and $p(t)$ is the ideal sinc pulse with unit energy, i.e., $p(t)=\sqrt{B}\sinc(Bt)$. When the complex baseband signal is lowpass filtered at the receiver and sampled at the symbol rate, the resulting received signal sequence $\{y[l]\}$ in the complex baseband becomes (see \cite[Ch.~9]{bjornson2024introduction} for details)
\begin{align}
y[l] = \sum_{\ell=0}^{T} h_{\vect{\Psi}}[\ell]  x[l-\ell]  + n[l],  \label{eq:symbol-sampled-model-repeated}
\end{align}
where $n[l]\sim\CN(0,N_0)$ is the filtered and sampled noise sequence. The channel is represented by the $T+1$ discrete-time channel coefficients $h_{\vect{\Psi}}[0],\ldots,h_{\vect{\Psi}}[T]$, where $T$ is determined by the delay spread, PAM pulse, and the bandwidth so that we have  $h_{\vect{\Psi}}[\ell]\approx 0$ for $\ell>T$. The channel coefficients are computed based on the end-to-end channel model in \eqref{eq:end-to-end-impulse} as
\begin{align} \nonumber
h_{\vect{\Psi}}[\ell] &= (p * g_{\vect{\Psi}} * p) \left(  t \right) \Big|_{t=\ell/B} =  (p * g_{\textrm{s}} * p) (t) \Big|_{t=\ell/B}  \nonumber \\
 & + \sum_{n=1}^{N}\sum_{m=1}^{N} (p * g_{\textrm{r},n} * \vartheta_{n,m; \vect{\Psi}} * g_{\textrm{t},m} * p) (t) \Big|_{t=\ell/B}.  \label{eq:end-to-end-impulse-response-discrete}
\end{align}
 In the above equation, the function $p(t)$ has dual purposes. It acts as both the pulse shape for PAM and the complex baseband equivalent ideal lowpass filter.

We assume the connection between any pairs of RIS elements acts as a linear phase filter within the signal band, consistent with \cite{9721205,li2024wideband} and good RIS design principles.
Hence, its passband impulse response can be expressed as $ \vartheta_{n,m,\pb; \vect{\Psi}} (t) = \gamma_{n,m}\delta(t  - \tau_{\psi_{n,m}})$, where $\gamma_{n,m}$ is the amplitude loss and $\tau_{\psi_{n,m}}$ is the delay experienced from RIS element $n$ to RIS element $m$, which results in the downshifted filter
\begin{equation} \label{eq:basic-RIS-impulse-response}
\vartheta_{n,m,\pb; \vect{\Psi}} (t) = \gamma_{n,m}\delta(t  - \tau_{\psi_{n,m}}) e^{-\imagunit 2\pi f_{\mathrm{c}} t},
 \end{equation}
where $\gamma_{n,m}$ and $\tau_{\psi_n}$ are the controllable parameters.

To model general wideband channels, we assume the static channel's impulse response consists of 
 $L_{\textrm{s}}$ propagation paths:
\begin{align} \label{eq:passband-channel-model-static}
 \quad g_{\textrm{s}}(t) = \sum_{i=1}^{L_{\textrm{s}}} \alpha_{\textrm{s},i}  e^{-\imagunit 2\pi f_{\mathrm{c}} t} \delta( t + \eta - \tau_{\textrm{s},i}),
\end{align}
where $\alpha_{\textrm{s},i} \in [0,1]$ represents the attenuation and $\tau_{\textrm{s},i} \geq 0$ represents the delay of path $i$, where $i=1,\ldots,L_{\textrm{s}}$.
The $\eta$ denotes the receiver's clock delay, which must be selected as described in \cite[Sec.~7.1]{bjornson2024introduction} to obtain a causal filter representation.

We define the number of propagation paths between the transmitter and the RIS as $L_{\textrm{t}}$ and between the surface and the receiver as $L_{\textrm{r}}$. Similar to \eqref{eq:passband-channel-model-static}, we can model the impulse responses to and from RIS element $n$ as follows:
\begin{align}
g_{\textrm{t},n}(t) &= \sum_{i=1}^{L_{\textrm{t}}} \alpha_{\textrm{t},n,i} e^{-j 2\pi f_{\textrm{c}} t} \delta(t - \tau_{\textrm{t},n,i}), \label{eq:passband-channel-model-t-atom-n}\\
g_{\textrm{r},n}(t) &= \sum_{j=1}^{L_{\textrm{r}}} \alpha_{\textrm{r},n,j} e^{-j 2\pi f_{\textrm{c}} t} \delta(t + \eta - \tau_{\textrm{r},n,j}), \label{eq:passband-channel-model-r-atom-n}
\end{align}
where $\alpha_{\textrm{t},n,i},\alpha_{\textrm{r},n,j} \in [0,1]$ represent the attenuations and $\tau_{\textrm{t},n,i},\tau_{\textrm{r},n,j} \geq 0$ represent the propagation delays.

Substituting the RIS impulse responses from \eqref{eq:basic-RIS-impulse-response} and the channel impulse responses from \eqref{eq:passband-channel-model-static}--\eqref{eq:passband-channel-model-r-atom-n} into \eqref{eq:end-to-end-impulse-response-discrete}, the channel coefficients can be calculated as
\begin{align} \nonumber
&h_{\boldsymbol{\psi}}[\ell] =  \sum_{i=1}^{L_{\textrm{s}}}  \alpha_{\textrm{s},i}  e^{-\imagunit 2\pi f_{\mathrm{c}} (\tau_{\textrm{s},i}-\eta)}
\sinc \big( \ell + B( \eta -\tau_{\textrm{s},i}) \big) +\\
& \sum_{n=1}^{N}   \sum_{m=1}^{N} \sum_{j=1}^{L_{\textrm{r}}} \sum_{i=1}^{L_{\textrm{t}}} \alpha_{\textrm{r},n,j}  \alpha_{\textrm{t},m,i} \gamma_{n,m} e^{-\imagunit 2\pi f_{\mathrm{c}} (\tau_{\textrm{r},n,j}+\tau_{\textrm{t},m,i}+\tau_{\psi_{n,m}} - \eta)}
\nonumber\\
&\times\sinc \big( \ell + B( \eta -\tau_{\textrm{r},n,j}-\tau_{\textrm{t},m,i}) \big)
 \label{eq:example-end-to-end}
\end{align}
for $\ell=0,\ldots,T$, using the properties $(p * p)(t) = \sinc(Bt)$ and the convolution between $\sinc(Bt)$ and $e^{-\imagunit 2\pi f_{\mathrm{c}} t} \delta(t- \tau)$ being $ \sinc(B(t-\tau)) e^{-\imagunit 2\pi f_{\mathrm{c}} \tau}$. The last sinc term in \eqref{eq:example-end-to-end}, which is originally $\sinc ( \ell + B( \eta -\tau_{\textrm{r},n,j}-\tau_{\textrm{t},m,i}-\tau_{\psi_{n,m}}))$, is simplified by ignoring the term containing $\tau_{\psi_{n,m}}$ since the delay caused by the reflection is negligible compared to the symbol time $1/B$, given that $\tau_{\psi_{n,m}} /(1/B) = B \tau_{\psi_{n,m}} \approx 0$. However, the RIS introduces a noticeable phase-shift since $f_{\mathrm{c}} \gg B$.

When there are far-field conditions between the transmitter/receiver/scatterers and the RIS, each path gives rise to a constant attenuation across the RIS: $\alpha_{\textrm{t},n,i} = \alpha_{\textrm{t},1,i}$ and $\alpha_{\textrm{r},n,j} = \alpha_{\textrm{r},1,j}$ for all $n$, where we use the first RIS element taken as the reference. Then, we can associate the $i$th incident path to the surface with an angle pair $(\varphi_{\textrm{i},i},\theta_{\textrm{i},i})$, measured from the broadside direction of the RIS, and the $j$th outgoing path with an angle pair $(\varphi_{\textrm{o},j},\theta_{\textrm{o},j})$. Gathering the $N$ phase-shifts related to such a path, we can express them using the following array response vectors:
\begin{align}
\vect{a}(\varphi_{\textrm{i},i},\theta_{\textrm{i},i}) &=  \begin{bmatrix}
1 \\
e^{-\imagunit 2\pi f_{\mathrm{c}} (\tau_{\textrm{t},2,i} - \tau_{\textrm{t},1,i} )} \\
\vdots \\
e^{-\imagunit 2\pi f_{\mathrm{c}} (\tau_{\textrm{t},N,i} - \tau_{\textrm{t},1,i} )}
\end{bmatrix}, \\
 \vect{a}(\varphi_{\textrm{o},j},\theta_{\textrm{o},j}) &= \begin{bmatrix}
1 \\
e^{-\imagunit 2\pi f_{\mathrm{c}} (\tau_{\textrm{r},2,j} - \tau_{\textrm{r},1,j} )} \\
\vdots \\
e^{-\imagunit 2\pi f_{\mathrm{c}} (\tau_{\textrm{r},N,j} - \tau_{\textrm{r},1,j} )}
\end{bmatrix}.
\end{align}
For a given path, the delay variations across the surface are negligible, such that $B\tau_{\textrm{r},n,j} \approx B\tau_{\textrm{r},1,j}$ and $B\tau_{\textrm{t},n,i} \approx B\tau_{\textrm{t},1,i}$ for all $n$. We can utilize these properties to rewrite the channel coefficients in \eqref{eq:example-end-to-end} as
\begin{align}
h_{\boldsymbol{\psi}}[\ell] = c_{\textrm{s}}[\ell] + 
\sum_{j=1}^{L_{\textrm{r}}} \sum_{i=1}^{L_{\textrm{t}}}  
c_{i,j}[\ell]
 \vect{a}^{\Ttran}(\varphi_{\textrm{o},j},\theta_{\textrm{o},j})  \vect{\Psi} 
\vect{a}(\varphi_{\textrm{i},i},\theta_{\textrm{i},i}),
 \label{eq:example-end-to-end2}
\end{align}
where the reflection matrix $\vect{\Psi}$ contains the controllable amplitudes and phase-shifts induced between each pair of RIS elements. Its $(n,m)$th entry is given as $[\vect{\Psi}]_{n,m}=\gamma_{n,m}e^{-\imagunit 2\pi f_{\mathrm{c}} \tau_{\psi_{n,m}}}$ and the other terms are 
\begin{align}
c_{\textrm{s}}[\ell]  &=   \sum_{i=1}^{L_{\textrm{s}}}  \alpha_{\textrm{s},i}  e^{-\imagunit 2\pi f_{\mathrm{c}} (\tau_{\textrm{s},i}-\eta)}
\sinc \big( \ell + B( \eta -\tau_{\textrm{s},i}) \big),  \\
c_{i,j}[\ell] &=   \alpha_{\textrm{r},1,j} \alpha_{\textrm{t},1,i}  e^{-\imagunit 2\pi f_{\mathrm{c}} (\tau_{\textrm{r},1,j}+\tau_{\textrm{t},1,i}- \eta)} \nonumber \\
&\quad \times \sinc \big( \ell + B( \eta -\tau_{\textrm{r},1,j}-\tau_{\textrm{t},1,i}) \big).
\end{align}

Building on \cite{shen2021modeling,zhou2023optimizing}, a lossless and reciprocal RIS with a fully-connected BD-RIS circuitry tuned to maximize the reflected power should exhibit a symmetric and unitary $\vect{\Psi}$:
\begin{align}
    \vect{\Psi}=\vect{\Psi}^{\Ttran}, \quad \vect{\Psi}\vect{\Psi}^{\Htran} = \vect{I}_N. \label{eq:constraints}
\end{align}
If we apply OFDM modulation with $S$ data symbols per block and a cyclic prefix\index{cyclic prefix} of length $T$ (with $S>T$), it follows from \cite[Sec.~7.1.1]{bjornson2024introduction} that we obtain $S$ memoryless subcarriers of the kind
\begin{align} 
\bar{y}[\nu] =\bar{h}_{\boldsymbol{\psi}}[\nu] \bar{\chi}[\nu] + \bar{n}[\nu], \quad \textrm{for} \,\, \nu = 0, \ldots, S-1, \label{eq:symbol-sampled-model-OFDM-repeat2}
\end{align}
where $\bar{n}[\nu]\sim \CN(0,N_0)$ is the independent noise and the reconfigurable frequency-domain channel coefficients are
\begin{align}
\bar{h}_{\boldsymbol{\psi}}[\nu] = \bar{c}_{\textrm{s}}[\nu] + 
\sum_{j=1}^{L_{\textrm{r}}} \sum_{i=1}^{L_{\textrm{t}}}  
\bar{c}_{i,j}[\nu]
 \vect{a}^{\Ttran}(\varphi_{\textrm{o},j},\theta_{\textrm{o},j})  \vect{\Psi} 
\vect{a}(\varphi_{\textrm{i},i},\theta_{\textrm{i},i})
 \label{eq:subcarrier-RIS}
\end{align}
where
\begin{align} \label{}
\bar{c}_{\textrm{s}}[\nu] &= \sum_{\ell = 0}^{T} c_{\textrm{s}}[\ell]  e^{-\imaginary 2 \pi \ell \nu /S}, \quad  \nu = 0,\ldots,S-1, \\
\bar{c}_{i,j}[\nu] &= \sum_{\ell = 0}^{T} c_{i,j}[\ell] e^{-\imaginary 2 \pi \ell \nu /S}, \quad  \nu = 0,\ldots,S-1.
\end{align}
The expression in \eqref{eq:subcarrier-RIS} can be expressed as 
\begin{align}
&\bar{h}_{\boldsymbol{\psi}}[\nu]= \underbrace{\bar{c}_{\textrm{s}}[\nu]}_{\bar{h}_{\nu}}\nonumber\\
&+\tr\Bigg(\vect{\Psi}\underbrace{\sum_{j=1}^{L_{\textrm{r}}} \sum_{i=1}^{L_{\textrm{t}}}  
\bar{c}_{i,j}[\nu]
 \vect{a}(\varphi_{\textrm{i},i},\theta_{\textrm{i},i})\vect{a}^{\Ttran}(\varphi_{\textrm{o},j},\theta_{\textrm{o},j})}_{=\vect{H}_\nu}  \Bigg).  \label{eq:subcarrier-RIS-rewrite}
\end{align} 
Interestingly, this rigorously derived OFDM system model differs from the one proposed in \cite{li2024wideband}. A closer comparison cannot be made since that paper does not disclose all details.

\section{Wideband Capacity Maximization}

Similar to \cite[Th.~7.2]{bjornson2024introduction}, we can write the capacity of the considered BD-RIS-assisted single-input single-output (SISO) OFDM system for a given reflection matrix $\vect{\Psi}$ as
\begin{equation} \label{eq:capacity-wideband-RIS}
C =\frac{B}{T+S}   \sum_{\nu=0}^{S-1} \log_2 \left( 1 + \frac{q_{\nu}^{\mathrm{opt}}  \left| \bar{h}_\nu+\tr\left(\vect{\Psi}\vect{H}_{\nu}\right) \right|^2}{N_0}  \right) \quad \textrm{bit/s},
\end{equation}
where $B$ is the bandwidth,
\begin{equation} \label{eq:q_nu_star-repeated-RIS}
q_{\nu}^{\mathrm{opt}} = 
 \max \left( \mu - \frac{N_0}{\left| \bar{h}_\nu+\tr\left(\vect{\Psi}\vect{H}_{\nu}\right) \right|^2}, 0 \right), \,\,\, \nu=0,\ldots,S-1,
\end{equation}
is the power assigned to subcarrier $\nu$ using water-filling,
and the variable $\mu$ is selected to satisfy the total power constraint $\sum_{\nu=0}^{S-1} q_{\nu}^{\mathrm{opt}} = q S$. Note that $q$ denotes the total power.

We want to maximize the capacity in \eqref{eq:capacity-wideband-RIS} with respect to $\vect{\Psi}$, subject to the constraints on the reflection matrix given in \eqref{eq:constraints}. This is a challenging optimization problem since $\vect{\Psi}$ affects all $S$ subcarriers. Instead of addressing this difficult problem as a whole, we will adopt a divide-and-conquer approach.

First, we will follow the low-complexity \emph{power method} outlined in \cite[Ch.~9]{bjornson2024introduction} and maximize the total channel gain over all subcarriers. This is a simpler objective compared to directly dealing with the capacity expression, which involves water-filling-based power coefficients. The metric can be motivated by the fact that the low-SNR approximation of the rate with equal power allocation is proportional to the total channel gain. Once we have obtained an RIS configuration using this approach, we will apply water-filling power allocation to the system. The total channel gain over all the subcarrier is
\begin{align} \label{eq:power-subcarriers-RIS}
& \sum_{\nu=0}^{S-1} \left| \bar{h}_{\boldsymbol{\psi}}[\nu] \right|^2 =  \sum_{\nu=0}^{S-1} \left| \bar{h}_{\nu} + \vect{h}_{\nu}^{\Ttran}\boldsymbol{\psi}  \right|^2 =\sum_{\nu=0}^{S-1} \left|\bar{h}_{\nu} \right|^2 \nonumber \\
&+ \boldsymbol{\psi}^{\Htran} \underbrace{\left( \sum_{\nu=0}^{S-1}\vect{h}^*_\nu \vect{h}^{\Ttran}_\nu \right)}_{= \vect{A}} \boldsymbol{\psi}+2\Re\Bigg(\boldsymbol{\psi}^{\Htran}\underbrace{\left( \sum_{\nu=0}^{S-1}\bar{h}_\nu \vect{h}^*_\nu  \right)}_{= \vect{b}}\Bigg),
\end{align}
where $\vect{h}_{\nu}=\mathrm{vec} \left(\vect{H}_{\nu}\right)$ and $\boldsymbol{\psi}=\mathrm{vec}(\vect{\Psi})$. We will use notation $\vect{A}, \vect{b}$ defined above for notational convenience in this section.

Relaxing the constraints in \eqref{eq:constraints} into $\Vert \boldsymbol{\psi}\Vert^2= N$, we propose to solve the problem
\begin{subequations} \label{eq:optimization3}
\begin{align} 
&\underset{\boldsymbol{\psi}}{\textrm{maximize}} \quad 
\boldsymbol{\psi}^{\Htran}\vect{A}\boldsymbol{\psi}+2\Re\left(\boldsymbol{\psi}^{\Htran}\vect{b}\right)
 \label{eq:norm-maximization5} \\
&\textrm{subject to} \quad \Vert \boldsymbol{\psi}\Vert^2= N. \label{eq:norm-maximization6}\end{align}
\end{subequations}
This problem has the same form as in \cite[Lem.~1]{Demir2021RIS}, thus, its solution can be found by the following approach.
Let $\lambda_{d}\geq0$ be the nonnegative eigenvalues and $\vect{u}_{d}\in \mathbb{C}^{N^2}$ be the corresponding orthonormal eigenvectors of $\vect{A}$, for $d=1,\ldots,N^2$. If $\vect{b}=\vect{0}$, the optimal solution to the problem \eqref{eq:optimization3} is given by $\boldsymbol{\psi}^{\star}=\sqrt{N}\vect{u}_{\bar{d}}$, where $\bar{d}$ is the index corresponding to the dominant eigenvector. Otherwise, the optimal solution is  
\begin{equation}
\boldsymbol{\psi}^{\star} = \sum_{d=1}^{N^2} \frac{\vect{u}_{d}\vect{u}_{d}^{\Htran}\vect{b}}{\gamma^{\star}-\lambda_{d}}, \quad d=1,\ldots,N , \label{eq:phi}
\end{equation}
where $\gamma^{\star}>\max_d\lambda_{d}$ is the unique root of$\sum_{d=1}^{N^2}\frac{\left\vert\vect{u}_{d}^{\Htran}\vect{b}\right\vert^2}{\left(\gamma-\lambda_{d}\right)^2}=N$.
\begin{algorithm}[t!]
	\caption{Wideband capacity maximization for BD-RIS-assisted SISO-OFDM system.} \label{alg:proposed-method}
	\begin{algorithmic}[1]
 \State Construct $\vect{A}$ and $\vect{b}$ as in \eqref{eq:power-subcarriers-RIS}
 \State Obtain the optimal solution to \eqref{eq:optimization3} and denote it by  $\boldsymbol{\psi}^{\star}$
 \State  $\overline{\vect{\Psi}} \leftarrow \mathrm{vec}^{-1}(\boldsymbol{\psi}^{\star})$
 \State Obtain the singular value decomposition of $\frac{\overline{\vect{\Psi}}+\overline{\vect{\Psi}}^{\Ttran}}{2}$ as $\vect{S}\vect{\Sigma}\vect{S}^{\Ttran}$ 
 \State $\vect{\Psi}^{\star} \leftarrow \vect{S}\vect{S}^{\Ttran}$ 
 \State Construct $\overline{\vect{A}}$ as in \eqref{eq:power-subcarriers-RIS2}
		\State 
  Select $\vect{d}_0 = [1,\ldots,1]^{\Ttran}$ and the number of iterations $L$
		\For{$i=0,\ldots,L-1$} 
		\State $\vect{w}_{i+1} \gets \frac{\overline{\vect{A}}\vect{d}_{i}}{\| \overline{\vect{A}}\vect{d}_{i}\|}$
		\State $\phi \gets \arg\left([\vect{w}_{i+1}]_{1}\right)$
		\State $\vect{d}_{i+1} \leftarrow 
  \left [1,e^{\imaginary (\arg\left([\vect{w}_{i+1}]_{2}\right)-\phi)},\ldots,e^{\imaginary (\arg\left([\vect{w}_{i+1}]_{N+1}\right)-\phi)}\right]^{\Ttran}$
		\EndFor
  \State $\vect{D} \leftarrow \diag\left(e^{\imaginary (\arg\left([\vect{d}_{i+1}]_{2}\right)},\ldots, e^{\imaginary (\arg\left([\vect{d}_{i+1}]_{N+1}\right)}\right)$
  \State $\vect{\Psi}\leftarrow \vect{S}\vect{D}\vect{S}^{\Ttran}$
  \State Apply the water-filling power allocation and obtain $q_{\nu}^{\rm opt}$, for $\nu=0,\ldots,S-1$.
		\State {\bf Output:} $\vect{\Psi}$, $q_{\nu}^{\rm opt}$ from \eqref{eq:q_nu_star-repeated-RIS}, for $\nu=0,\ldots,S-1$
	\end{algorithmic}
\end{algorithm}
After getting the optimal vector $\boldsymbol{\psi}^{\star}$, we reshape it into a $N\times N$ matrix as $\overline{\vect{\Psi}} = \mathrm{vec}^{-1}(\boldsymbol{\psi}^{\star})$ and our next step is to find the nearest $\vect{\Psi}$ to $\overline{\vect{\Psi}}$, which satisfy the original constraints in \eqref{eq:constraints}. The corresponding optimization problem is
\begin{subequations} \label{eq:optimization4}
\begin{align} 
&\underset{\vect{\Psi}}{\textrm{maximize}} \quad 
\left\Vert \vect{\Psi}-\overline{\vect{\Psi}} \right\Vert_{\rm F}
 \label{eq:norm-minimization} \\
&\textrm{subject to} \quad \vect{\Psi} = \vect{\Psi}^{\Ttran}, \quad \vect{\Psi}\vect{\Psi}^{\Htran}=\vect{I}_N. \label{eq:constraints2}\end{align}
\end{subequations}
The solution to the problem \eqref{eq:optimization4} is given in Theorem~\ref{theorem1}.
\begin{theorem} \label{theorem1} Let $\vect{S}\vect{\Sigma}\vect{S}^{\Ttran}$ denote the singular value decomposition of $\frac{\overline{\vect{\Psi}}+\overline{\vect{\Psi}}^{\Ttran}}{2}$. The optimal solution to \eqref{eq:optimization4} is 
\begin{align}
    \vect{\Psi}^{\star} = \vect{S}\vect{S}^{\Ttran}. 
\end{align}
\end{theorem}
\vspace{-4mm}
\begin{proof}
Since $\vect{\Psi}$ is a symmetric and unitary matrix according to the constraints in \eqref{eq:constraints}, it can be decomposed as $\vect{\Psi}=\vect{U}\vect{U}^{\Ttran}$ for some unitary matrix $\vect{U}\in \mathbb{C}^{N\times N}$. From the triangular inequality, it follows that
\begin{align}
&\left \Vert \vect{U}\vect{U}^{\Ttran}-\frac{\vect{Q}+\vect{Q}^{\Ttran}}{2} \right \Vert_{\rm F} \leq \frac{1}{2}\left \Vert \vect{U}\vect{U}^{\Ttran}-\vect{Q}\right\Vert_{\rm F}\nonumber\\
&+ \frac{1}{2}\left \Vert \vect{U}\vect{U}^{\Ttran}-\vect{Q}^{\Ttran}\right\Vert_{\rm F} = \left \Vert \vect{U}\vect{U}^{\Ttran}-\vect{Q}\right\Vert_{\rm F}.
\end{align}
Denoting the singular value decomposition of $\frac{\overline{\vect{\Psi}}+\overline{\vect{\Psi}}^{\Ttran}}{2}$ by $\vect{S}\vect{\Sigma}\vect{S}^{\Ttran}$, for any unitary matrix $\vect{U}$ we obtain the relation
\begin{align}
   \left \Vert \vect{U}\vect{U}^{\Ttran}-\vect{Q}\right\Vert_{\rm F} \geq \left \Vert \vect{U}\vect{U}^{\Ttran}-\vect{S}\vect{\Sigma}\vect{S}^{\Ttran}\right\Vert_{\rm F}. 
\end{align}
The right-hand side of the above inequality, as a function of $\vect{U}$, is minimized when $\vect{U}=\vect{S}$. Hence, for any arbitrary unitary matrix $\vect{U}$, it holds that
\begin{align}
   \left \Vert \vect{U}\vect{U}^{\Ttran}-\vect{Q}\right\Vert_{\rm F} \geq  \left \Vert \vect{S}\vect{S}^{\Ttran}-\vect{Q}\right\Vert_{\rm F}.
\end{align}
This means that $\vect{S}\vect{S}^{\Ttran}$ is the optimal solution to \eqref{eq:optimization4}.
\end{proof}

After getting the solution $ \vect{\Psi}^{\star} = \vect{S}\vect{S}^{\Ttran}$, we can refine $\vect{\Psi}$ by decomposing it $\vect{\Psi} = \vect{S}\vect{D}\vect{S}^{\Ttran}$, where $\vect{D} =\diag(e^{\imagunit\theta_1},\ldots,e^{\imagunit\theta_N})$ is a diagonal matrix of phase-shifts, which do not destroy the constraints in \eqref{eq:constraints2} and provide an additional refinement opportunity at the final step of the proposed method. Substituting $\vect{\Psi} = \vect{S}\vect{D}\vect{S}^{\Ttran}$ into the total channel gain $\sum_{\nu=0}^{S-1} \left| \bar{h}_{\boldsymbol{\psi}}[\nu] \right|^2 $ and denoting $\vect{d}=[1 \ \mathrm{vec}^{\Ttran}(\vect{D})]^{\Ttran}\in \mathbb{C}^{N+1}$, we obtain \begin{align} \label{eq:power-subcarriers-RIS2}
& \sum_{\nu=0}^{S-1} \left| \bar{h}_{\boldsymbol{\psi}}[\nu] \right|^2 = \vect{d}^{\Htran}\underbrace{\sum_{\nu=0}^{S-1}\vect{f}_{\nu}\vect{f}_{\nu}^{\Htran}}_{\overline{\vect{A}}}\vect{d},
\end{align}  
where
\begin{align}
   \vect{f}_\nu = \begin{bmatrix}
   \bar{h}_{\nu} \\ \diag\left( \vect{S}^{\Ttran}\vect{H}_{\nu}\vect{S} \right)
   \end{bmatrix}.
\end{align}
To maximize $\vect{d}^{\Htran}\overline{\vect{A}}\vect{d}$ under the constraints $[\vect{d}]_1=1$ and $\left\vert [\vect{d}]_n \right\vert =1$, for $n=2,\ldots,N+1$, the power iteration method in \cite[Ch.~9]{bjornson2024introduction} can be applied. The overall procedure of the proposed method is outlined in Algorithm~\ref{alg:proposed-method}. The main complexity of the algorithm arises from obtaining the singular value decomposition of an $N^2 \times N^2$ matrix. The numerical experiments provided in the next section had an algorithmic runtime less than one second on a personal computer for $N=64$ RIS elements and $S=2000$ subcariers.

\section{Numerical Results}

In this section, we compare the capacities obtained over a BD-RIS-assisted SISO-OFDM channel with various configuration schemes. We also consider a conventional RIS with a diagonal $\vect{\Psi}$-matrix that is optimized using the power iteration method described in \cite[Alg.~9.1]{bjornson2024introduction}. For the BD-RIS, we compare our proposed Algorithm~\ref{alg:proposed-method} with a benchmark that optimizes the reflection matrix for the principal rank-one component of each channel tap using the method described in \cite{nerini2023closed}, and selects the configuration that yields the highest power across the taps. This benchmark is referred to as ``strongest tap maximization.'' We focus on the principal rank-one component because the closed-form optimization presented in \cite{nerini2023closed} is specifically tailored for that structure. In addition, we examine a random configuration scheme for the BD-RIS, wherein the unitary symmetric matrix $\vect{S}$ is randomly generated and a power iteration method is employed to adjust the diagonal matrix $\vect{D}$.

The RIS is positioned in the $yz$-plane, facing the positive $x$-axis, with its center at $(0,0,0)$ meters. The transmitter and receiver are located at $(40,-40,0)$ meters and $(20,0,0)$ meters, respectively. We assume LOS channels with multiple reflected paths to and from the RIS, while the static channel is NLOS when it exists. The carrier frequency is $3$ GHz and the RIS comprises $64$ elements arranged in an $8\times 8$ grid, with each element having a size of $\lambda/4 \times \lambda/4$. The channel modeling follows the 3GPP channel model guidelines in \cite{3GPP25996}. The capacity is calculated as the average over random realizations of the multipath components' characteristics. We will present the capacity as a function of the bandwidth $B$, assuming that the transmit power scales proportionally with the bandwidth to maintain a signal power spectral density of $1$ W per MHz. The subcarrier spacing is $150$ kHz, leading to an increase in the number of subcarriers and channel taps with the bandwidth, but the multipath environment is unchanged.

\begin{figure}[t!]
        \centering
	\begin{overpic}[width=.9\columnwidth,tics=10]{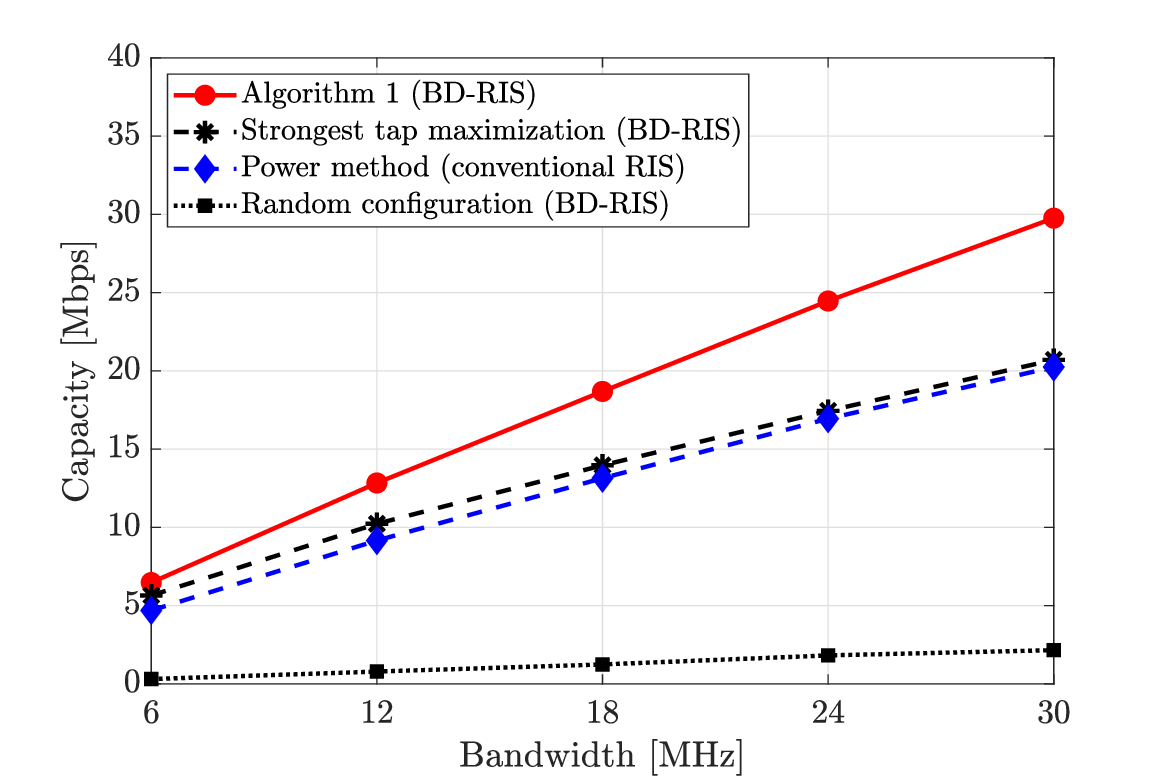}
\end{overpic} 
\vspace{-4mm}
        \caption{The capacity achieved for differently configured surfaces without a static path and with NLOS channels between the transmitter/receiver and RIS. }
        \label{fig:1}
        \vspace{-4mm}
\end{figure}

In Fig.~\ref{fig:1}, we consider a scenario without a static channel and NLOS transmitter-RIS and RIS-receiver channels without any LOS paths. As shown in the figure, optimizing the RIS for BD operation is crucial, and our proposed algorithm significantly outperforms the benchmark methods. It provides up to a 50\% improvement in capacity compared to both the strongest path maximization approach and the conventional diagonal RIS.

\begin{figure}[t!]
        \centering
	\begin{overpic}[width=.9\columnwidth,tics=10]{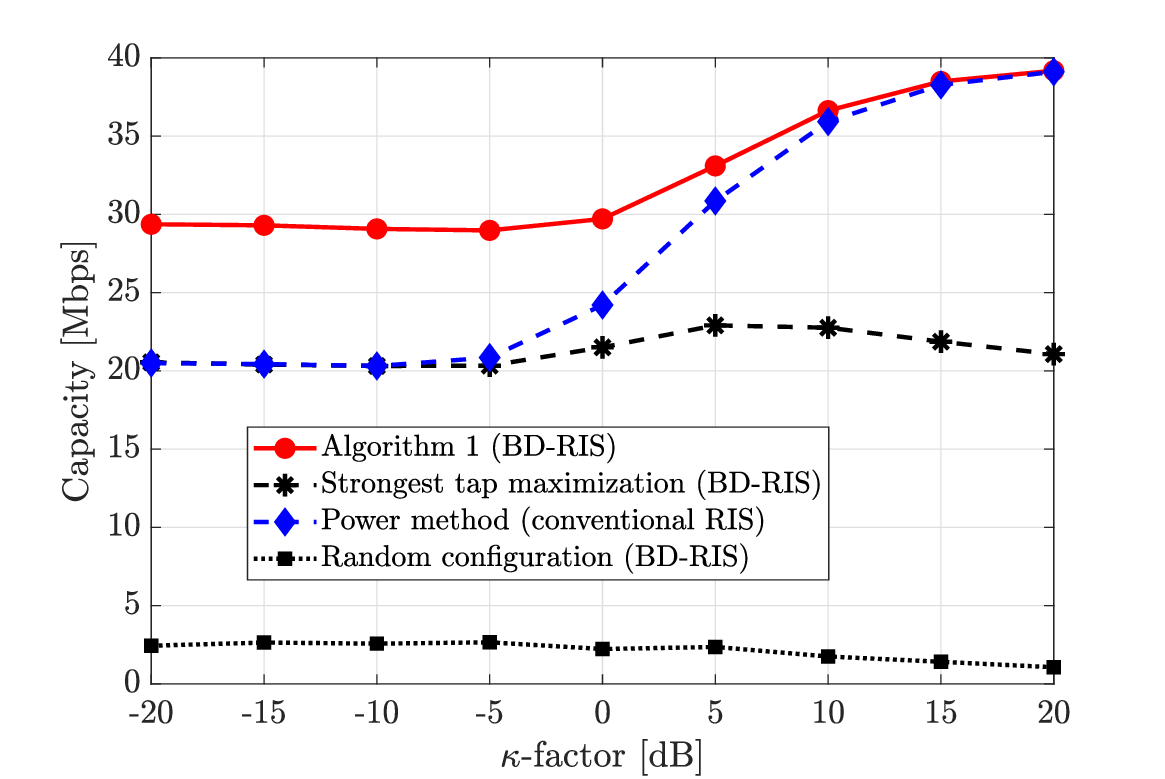}
\end{overpic} 
\vspace{-4mm}
        \caption{The capacity achieved for differently configured surfaces when the static channel is zero and the bandwidth is $B=30$\,MHz. }
        \label{fig:2}
        \vspace{-6mm}
\end{figure}

\begin{figure}[t!]
        \centering
	\begin{overpic}[width=.9\columnwidth,tics=10]{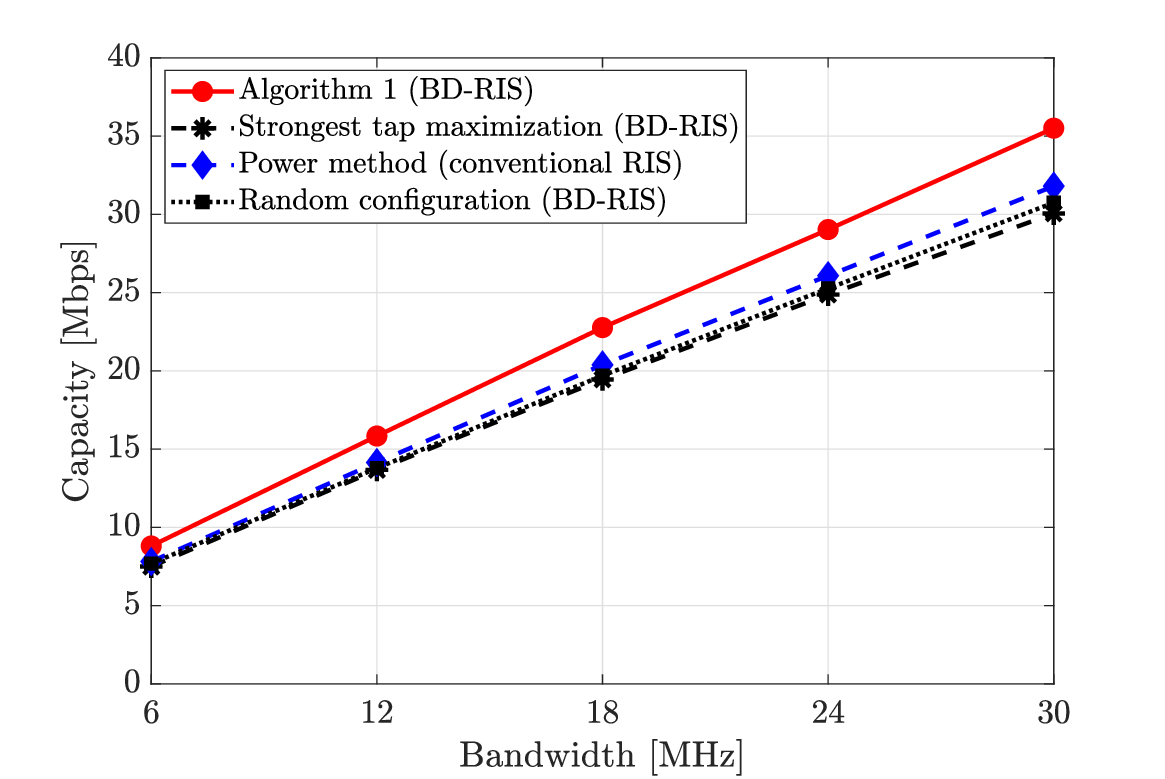}
\end{overpic}
\vspace{-4mm}
        \caption{The capacity achieved for differently configured surfaces when the static channel is non-zero and the $\kappa$-factor of the channels is zero. }
        \label{fig:3}
        \vspace{-6mm}
\end{figure}

In Fig.~\ref{fig:2}, we set the bandwidth to $B = 30$\,MHz and consider the presence of a LOS path between the transmitter/receiver and the RIS. We vary the $\kappa$-factor of the resulting Rician fading. As the $\kappa$-factor increases (i.e., as the LOS paths become more dominant), the capacity benefit of the BD-RIS operation decreases; it disappears for $\kappa>10$. Finally, Fig.~\ref{fig:3} introduces a static channel with a channel gain that is 40\,dB weaker than the one described in \cite{3GPP25996}. Despite its lower gain, the random configuration performs similarly to the optimized diagonal RIS. In conclusion, the presence of dominant LOS paths and/or a static channel reduces the performance improvement of BD-RIS relative to conventional diagonal RIS.

\section{Conclusions}

In this letter, we have derived the SISO-OFDM system model for BD-RIS-assisted communications. The model and methodology are essential for future research since practical systems feature wideband channels.
We presented the first capacity expression for this setup and proposed an efficient algorithm for improving it with respect to the non-diagonal RIS reflection matrix by maximizing the total channel gain.
This algorithm outperforms traditional benchmarks.
Notably, replacing conventional RIS with BD-RIS enhances capacity in wideband scenarios, and the relative performance gap remains constant as bandwidth increases. However, the presence of static paths and/or LOS paths reduces the performance gains.

\bibliographystyle{IEEEtran}
\bibliography{IEEEabrv,refs}

\begin{thebibliography}{10}
\providecommand{\url}[1]{#1}
\csname url@samestyle\endcsname
\providecommand{\newblock}{\relax}
\providecommand{\bibinfo}[2]{#2}
\providecommand{\BIBentrySTDinterwordspacing}{\spaceskip=0pt\relax}
\providecommand{\BIBentryALTinterwordstretchfactor}{4}
\providecommand{\BIBentryALTinterwordspacing}{\spaceskip=\fontdimen2\font plus
\BIBentryALTinterwordstretchfactor\fontdimen3\font minus
  \fontdimen4\font\relax}
\providecommand{\BIBforeignlanguage}[2]{{%
\expandafter\ifx\csname l@#1\endcsname\relax
\typeout{** WARNING: IEEEtran.bst: No hyphenation pattern has been}%
\typeout{** loaded for the language `#1'. Using the pattern for}%
\typeout{** the default language instead.}%
\else
\language=\csname l@#1\endcsname
\fi
#2}}
\providecommand{\BIBdecl}{\relax}
\BIBdecl

\bibitem{10316535}
H.~Li, S.~Shen, M.~Nerini, and B.~Clerckx, ``Reconfigurable intelligent
  surfaces 2.0: Beyond diagonal phase shift matrices,'' \emph{IEEE
  Communications Magazine}, vol.~62, no.~3, pp. 102--108, 2024.

\bibitem{yang2020intelligent}
Y.~Yang, B.~Zheng, S.~Zhang, and R.~Zhang, ``Intelligent reflecting surface
  meets {OFDM}: Protocol design and rate maximization,'' \emph{IEEE
  Transactions on Communications}, vol.~68, no.~7, pp. 4522--4535, 2020.

\bibitem{sun20223d}
G.~Sun, R.~He, B.~Ai, Z.~Ma, P.~Li, Y.~Niu, J.~Ding, D.~Fei, and Z.~Zhong, ``A
  {3D} wideband channel model for ris-assisted {MIMO} communications,''
  \emph{IEEE Transactions on Vehicular Technology}, vol.~71, no.~8, pp.
  8016--8029, 2022.

\bibitem{9721205}
E.~Björnson, H.~Wymeersch, B.~Matthiesen, P.~Popovski, L.~Sanguinetti, and
  E.~de~Carvalho, ``Reconfigurable intelligent surfaces: A signal processing
  perspective with wireless applications,'' \emph{IEEE Signal Processing
  Magazine}, vol.~39, no.~2, pp. 135--158, 2022.

\bibitem{shen2021modeling}
S.~Shen, B.~Clerckx, and R.~Murch, ``Modeling and architecture design of
  reconfigurable intelligent surfaces using scattering parameter network
  analysis,'' \emph{IEEE Transactions on Wireless Communications}, vol.~21,
  no.~2, pp. 1229--1243, 2021.

\bibitem{zhou2023optimizing}
Y.~Zhou, Y.~Liu, H.~Li, Q.~Wu, S.~Shen, and B.~Clerckx, ``Optimizing power
  consumption, energy efficiency and sum-rate using beyond diagonal {RIS}—a
  unified approach,'' \emph{IEEE Trans. Wirel. Commun.}, 2023.

\bibitem{li2022beyond}
H.~Li, S.~Shen, and B.~Clerckx, ``Beyond diagonal reconfigurable intelligent
  surfaces: From transmitting and reflecting modes to single-, group-, and
  fully-connected architectures,'' \emph{IEEE Transactions on Wireless
  Communications}, vol.~22, no.~4, pp. 2311--2324, 2022.

\bibitem{nerini2023closed}
M.~Nerini, S.~Shen, and B.~Clerckx, ``Closed-form global optimization of beyond
  diagonal reconfigurable intelligent surfaces,'' \emph{IEEE Transactions on
  Wireless Communications}, vol.~23, no.~2, pp. 1037--1051, 2024.

\bibitem{li2024wideband}
H.~Li, M.~Nerini, S.~Shen, and B.~Clerckx, ``Wideband modeling and beamforming
  for beyond diagonal reconfigurable intelligent surfaces,'' \emph{arXiv
  preprint arXiv:2403.12893v1}, Mar. 2024.

\bibitem{bjornson2024introduction}
E.~Bj{\"o}rnson and {\"O}.~T. Demir, ``Introduction to multiple antenna
  communications and reconfigurable surfaces,'' \emph{Now Publishers, Inc.},
  2024.

\bibitem{Demir2021RIS}
{\"O}.~T. Demir and E.~Bj{\"o}rnson, ``Is channel estimation necessary to
  select phase-shifts for {RIS}-assisted massive {MIMO}?'' \emph{IEEE
  Transactions on Wireless Communications}, vol.~21, no.~11, pp. 9537--9552,
  2022.

\bibitem{3GPP25996}
\emph{Spatial channel model for Multiple Input Multiple Output {(MIMO)}
  simulations (Release 16)}.\hskip 1em plus 0.5em minus 0.4em\relax {3GPP} {TS}
  25.996, Jul. 2020.

\end{thebibliography}

\end{document}